\documentstyle[aps,prd,epsf,floats]{revtex} 

\draft

\begin{document}
\twocolumn[\hsize\textwidth\columnwidth\hsize\csname
@twocolumnfalse\endcsname 

\title{No Pulsar Kicks from Deformed Neutrinospheres}

\author{H.-Thomas Janka}
\address{Max-Planck-Institut f\"ur Astrophysik, 
Karl-Schwarzschild-Str.~1, D-85740 Garching, Germany}

\author{Georg~G.~Raffelt} 
\address{Max-Planck-Institut f\"ur Physik 
(Werner-Heisenberg-Institut), 
F\"ohringer Ring 6, 80805 M\"unchen, Germany} 

\date{August 10, 1998}

\maketitle
             
\begin{abstract}
In a supernova core, magnetic fields cause a directional variation of
the neutrino refractive index so that resonant flavor oscillations
would lead to a deformation of the ``neutrinosphere'' for, say,
$\tau$ neutrinos.  The associated anisotropic neutrino emission was
proposed as a possible origin of the observed pulsar proper motions.
We argue that this effect was vastly overestimated because the
variation of the temperature over the deformed neutrinosphere is not
an adequate measure for the anisotropy of neutrino emission.  The
neutrino flux is generated inside the neutron star core and is
transported through the atmosphere at a constant luminosity, forcing
the temperature gradient in the atmosphere to adjust to the inflow of
energy from below.  Therefore, no emission anisotropy is caused by a
deformation of the neutrinosphere to lowest order. An estimate of the
higher-order corrections must take into account the modified
atmospheric temperature profile in response to the deformation of the
neutrinosphere and the corresponding feedback on the core. We go
through this exercise in the framework of a simplified model which can
be solved analytically.
\end{abstract}


\vskip2.0pc]


\section{Introduction}
\label{sec-1}

After the supernova collapse of a massive star, neutrinos carry away
about 99\% of the gravitational binding energy $E_{\rm b}$ of the
nascent neutron star, taking with them a huge amount of momentum which
is of the order $10^{43}~{\rm g~cm~s}^{-1}\,(E_{\rm b}/3\times
10^{53}~{\rm erg})$.  An anisotropy of the neutrino emission as small
as 1\% would suffice to account for a neutron star recoil of about
$300~{\rm km~s}^{-1}$~\cite{kicks} and could thus explain the observed
space velocities of most pulsars~\cite{pulsars}. However, even such a
small asymmetry is difficult to explain.

Pulsars tend to have strong magnetic fields, leading to the
speculation that $B$-fields could be a natural agent to cause
asymmetric neutrino emission.  For some time it appeared as if for
realistic field strengths the induced polarization of the nucleon
spins, together with the parity-violating properties of the
neutrino-nucleon cross-sections, was enough to explain the observed
pulsar kicks~\cite{Horowitz}.  Later it was recognized that this
``cumulative parity violation effect'' was in violation of fundamental
symmetries required of the Boltzmann collision equation; a correct
derivation leads to a much reduced anisotropy~\cite{Lai}.

This observation, together with the impressive recent evidence for
neutrino oscillations, leads one to take seriously another more
indirect mechanism. The neutrino refractive index depends on the
direction of the neutrino momentum relative to ${\bf B}$. For suitable
conditions, resonant neutrino oscillations can occur between the
neutrinospheres of electron neutrinos and, say, $\tau$ neutrinos,
leading to a deformation of the effective $\nu_\tau$~sphere~\cite{Kus96},
although the required conditions for large neutron star kicks may be 
rather extreme~\cite{Qian97}. The $\tau$ neutrinos would thus
be emitted from regions of varying effective temperatures, and thus,
it was argued, would be emitted anisotropically. This idea was then
taken up in several papers with modified neutrino oscillation
scenarios~\cite{Kus97,waste,morewaste,alsowaste}.

Unfortunately, however, this elegant scenario and its variations 
also appear to be fundamentally flawed in at least two serious ways.

The first problem is caused by a common misunderstanding of the
meaning of the ``effective temperature'' of the neutrino flux emerging
from a supernova core. It is usually thought that the total energy
carried away by neutrinos from a SN~core is roughly equipartitioned
between the flavors, yet the heavy-flavor neutrinos (we usually take
$\nu_\tau$ as an example) have stiffer spectra, i.e.\ their {\it
spectral\/} temperatures tend to be much larger than those of
$\bar\nu_e$. Evidently, the neutrino luminosities are not given by the
Stefan-Boltzmann law in a naive way---we will discuss this issue in
some detail in Sec.~\ref{sec-2}. For the moment it suffices to observe
that in a situation of exact flavor equipartition, a spectral swap of
two flavors by oscillations would not change the energy flux, except
perhaps indirectly by a response of the thermal medium to the
supposedly different spectra.

The flavor equipartition of the energy flux need not be exact, and for
the sake of argument we may contemplate a situation where most of the
energy is carried by $\nu_e$ and~$\bar\nu_e$. If oscillations take
place outside of the $\nu_\tau$ sphere, the oscillated $\nu_e$'s could
escape from different depths according to the $B$-field deformed
resonance sphere, and thus with different temperatures. Even then one
will not achieve a large flux asymmetry because it is not justified to
calculate the expected flux from the local gas temperature along the
resonance surface.

All of the neutrino spectra formation and oscillation physics of the
present problem take place in the ``atmosphere'' of the protoneutron
star, the outer region where the density drops quickly from core
values around nuclear density to effectively ``zero''. The neutrino
fluxes, however, are determined in the core of the neutron star.  The
atmosphere has virtually no heat capacity relative to the core.
Therefore, after a short time, typically of the order of a few hundred
milliseconds at most, which is very short compared with the
Kelvin-Helmholtz neutrino cooling time of the nascent neutron star,
the neutrino luminosity is governed by the core emission and the
surface-near layers have reached a state where the temperature
gradient ensures that all energy streaming up from below is carried
outwards with a luminosity that is independent of the radial position.

Therefore, the second serious problem of the oscillation kick scenario
is that, to lowest order, a shift of the neutrinosphere will leave the
neutrino luminosity unchanged.

A residual anisotropy effect obtains because the neutrino flux is not
strictly fixed by the core alone; it depends on the temperature at the
core-atmosphere interface. This temperature, in turn, depends on the
atmosphere so that there is an indirect influence of the atmospheric
structure on the neutrino fluxes.  More precisely, the neutrino flux
determines the temperature gradient in the atmosphere, and the
atmosphere influences the temperature at the core-atmosphere
interface.  Without a self-consistent treatment of this sort there is
no pulsar kick at all, and the kick that one does obtain is a
higher-order effect.

In the following discussion we will elaborate our two arguments in
more depth. In Sec.~\ref{sec-2} we will explain the connection between
the Stefan-Boltzmann law and the neutrino luminosities of a neutron
star. We will stress the inadequacy of a simplistic application of the
$R^2T^4$ scaling of the luminosity when $T$ is the spectral
temperature. In Sec.~\ref{sec-3} we will construct a simple
self-consistent model in the so-called Eddington atmosphere
approximation~\cite{Schi82}.  Our model leads to an estimate of the
higher-order emission anisotropy from a changed neutrinosphere by
direction-dependent resonant neutrino flavor conversions.  Finally,
Sec.~\ref{sec-4} is given over to a discussion and summary of our
findings.


\section{Neutrino transport in nascent neutron stars}
\label{sec-2}

\subsection{Neutrino Fluxes and the Stefan-Boltzmann Law}
\label{sec-21}

We begin our more detailed discussion with a description of some
crucial aspects of neutrino transport in nascent neutron stars. The
picture thus developed will serve as background information for the
analytical model of Sec.~\ref{sec-3}. The most important insight to be
presently explained is that the neutrino flux emerging from a
supernova core is not trivially given by the Stefan-Boltzmann law; the
spectral temperature does not fix the flux, in contrast with true
blackbody radiation~\cite{Janka}. 

Lepton number is lost from the collapsed stellar core by the emission
of electron neutrinos while energy is emitted in neutrinos and
antineutrinos of all flavors. Electron neutrinos are produced
efficiently via the $\beta$-process $e^-+p\to n+\nu_e$ during the
first second after collapse because of the high electron chemical
potential, so that the deleptonization, in particular of the
surface-near layers, proceeds very fast. Most of the gravitational
binding energy of the neutron star is radiated away after the
collapsed stellar core has settled into the static, compact and hot
protoneutron star when neutrinos and antineutrinos of all flavors take
up approximately the same share of the total energy and are emitted
with very similar luminosities from the thermal bath of the core.

The heat capacity and lepton number reservoir of the dense core are
much larger than those of the less dense and much less massive
atmosphere above.  Roughly, core and atmosphere are discerned by the
rather flat density gradient in the former, in contrast to the steep
density decline in the latter. The density at the core-atmosphere
interface is time-dependent and is typically between $10^{13}$ and
$10^{14}~{\rm g~cm}^{-3}$.  Its small heat capacity and short neutrino
diffusion time imply that the atmosphere radiates away its binding
energy in less than a few hundred milliseconds, to be compared with
the neutrino diffusion timescale out of the core of a few seconds and
the typical energy-loss timescale of several ten seconds.

Thus, after a brief initial relaxation phase, the neutrino luminosity
of the nascent neutron star is governed by the energy loss from the
core; it reaches its surface value already below the core-atmosphere
interface. Throughout the atmosphere, the luminosity is independent of
the radial position if gravitational redshift is ignored. The
temperature and density profiles in the atmospheric layers adjust to
the neutrino energy flux coming from inside to ensure its transport to
the stellar surface under the constraint of hydrostatic
equilibrium. This is equivalent to the situation in ordinary stars
where the photon luminosity is produced in the nuclear burning zones
while the stellar mantle and envelope adopt a structure in accordance
with the transport of this energy to the photosphere.  Evidently, the
energy flux of neutrino or antineutrino species $\nu_i$ cannot be
given simply by the local gas temperature according to
$F_{\nu_i}\propto T^4$.

When thermal equilibrium between neutrinos and the stellar medium is
assumed, the energy flux in the diffusion approximation can be
expressed in terms of the atmospheric temperature gradient as
\begin{equation}
F_{\nu_i}=-D_{\nu_i}\,{4\pi\over (hc)^3}\,4{\cal F}_3(0)\,
(kT)^3\,{\partial (kT)\over \partial r}\,.
\label{eq-2}
\end{equation}
Here, $D_{\nu_i}$ is the diffusion coefficient, suitably averaged over
the neutrino spectrum, while $h$, $c$ and $k$ are the Planck constant,
the speed of light, and Boltzmann's constant, respectively.  Further,
\begin{equation}
{\cal F}_j(\eta_{\nu_i})\equiv\int_0^\infty{\rm d}x\,
{x^j\over1+\exp(x-\eta_{\nu_i})}\,,
\label{eq-3}
\end{equation}
where $\eta_{\nu_i}$ is the neutrino degeneracy parameter, i.e.\ the
chemical potential divided by the temperature.  In Eq.~(\ref{eq-2}) it
was assumed that the neutrino degeneracy parameter is zero,
$\eta_{\nu_i} = 0$, which then gives ${\cal F}_3(0)\approx 3! =
6$. This is always true for $\mu$ and $\tau$ neutrinos and is also a
good approximation for electron neutrinos and antineutrinos after the
deleptonization of the atmosphere.

Another expression for the neutrino energy flux can be obtained
by relating it to the neutrino energy density,
\begin{equation}
\varepsilon_{\nu_i}={4\pi\over (hc)^3}\,(kT)^4\,{\cal F}_3(0)\,, 
\label{eq-3.1}
\end{equation}
which yields
\begin{equation}
F_{\nu_i}=c\,\langle{\mu}\rangle_{E,\nu_i}\,
{4\pi\over (hc)^3}\,
(kT)^4\,{\cal F}_3(0)\,.
\label{eq-4}
\end{equation}
The factor $\langle{\mu}\rangle_{E,\nu_i}$ denotes the average
cosine of the angle of neutrino propagation relative to the radial
direction. It is calculated from the neutrino phase-space distribution
function $f_{\nu_i}(r,t,\mu,\epsilon)$ according to
\begin{equation}
\langle{\mu}\rangle_{j,\nu_i}\equiv{\int_{-1}^{+1}{\rm d}\mu\,\mu
\int_0^\infty{\rm d}\epsilon\,\epsilon^{2+j}
f_{\nu_i}(r,t,\mu,\epsilon)  \over 
\int_{-1}^{+1}{\rm d}\mu\, 
\int_0^\infty{\rm d}\epsilon\,\epsilon^{2+j}
f_{\nu_i}(r,t,\mu,\epsilon)}\,,
\label{eq-5}
\end{equation}
where $r$ is the radial position, $t$ time, and $\epsilon$ the
neutrino energy.  Choosing $j = 0$ gives us the ``mean angle
cosine'' for the neutrino number flux, $\langle{\mu}\rangle_{
N,\nu_i}$, while $j = 1$ yields the corresponding quantity for the
energy flux, $\langle{\mu}\rangle_{E,\nu_i}$.
Comparing Eqs.~(\ref{eq-2}) and (\ref{eq-4}) gives
\begin{equation}
\langle{\mu}\rangle_{E,\nu_i}=-{4D_{\nu_i}\over c\,T}\,
{\partial T\over \partial r}={4\over 3}\,
{\langle{\lambda}\rangle_{E,\nu_i}\over h_T}\,,
\label{eq-6}
\end{equation}
where $h_T=(\partial\ln T/\partial r)^{-1}$ is the temperature scale
height and $D_{\nu_i} = c\langle{\lambda}\rangle_{E,\nu_i}/3$ was used
for the diffusion constant, with the mean free path $\lambda_{\nu_i}$
a suitable spectral average for the energy flux of neutrino
species~$\nu_i$.

From Eq.~(\ref{eq-4}) together with Eq.~(\ref{eq-6}) one verifies that
$F_{\nu_i}\propto T^4$ is not the whole story. Instead, the flux
factor $\langle{\mu}\rangle_{E,\nu_i}$ can be significantly different
from the canonical value $1/4$ which represents the Stefan-Boltzmann
law. Put another way, the energy flux of ``thermal'' radiation is
characterized by two parameters, the spectral temperature and the mean
angle cosine which quantifies the deviation from an isotropic
phase-space occupation.

Equation~(\ref{eq-6}) reveals that $\langle{\mu}\rangle_{E,\nu_i}$
depends on the position in the atmosphere because
$\langle{\lambda}\rangle_{E,\nu_i}$ becomes smaller for higher
temperature and density. In the protoneutron star atmosphere the
neutrino luminosity $L_{\nu} = 4\pi r^2 F_{\nu}$
(for an individual neutrino type or for the sum of
neutrino and corresponding antineutrino) is fixed by the inflow from
the core region. The flux factor $\langle{\mu}\rangle_{E,\nu_i}$, on
the other hand, increases with radius (decreasing temperature) in
accordance with Eq.~(\ref{eq-4}).

\subsection{Neutrino Spheres}
\label{sec-22}

Another frequently misunderstood concept is that of a
``neutrinosphere''. We stress that actually for each type of neutrino
two different kinds of neutrinospheres are defined, the ``energy
sphere'' with radius $R_{E,\nu_i}$ and the ``transport sphere'' with
radius $R_{{\rm t},\nu_i}$. The latter is what many authors mean with
{\it the\/} neutrinosphere, i.e.\ the surface of ``last scattering''
at optical depth 2/3
which emits the neutrino flux. The energy sphere is where
energy-exchanging reactions freeze out while energy-conserving
collisions may still be important.  Of course, the concept of
well-defined neutrinospheres or -surfaces is always a simplification
of the real situation because the neutrino-matter interactions are
strongly energy dependent.

When $R_{L,\nu_i}$ is the radius at which the luminosity $L_{\nu_i}$
has reached its surface value, the three radii obey the relation
$R_{L,\nu_i} < R_{E,\nu_i} < R_{{\rm t},\nu_i}$, because the
luminosity is fixed already deep inside the nascent neutron star.
Between $R_{L,\nu_i}$ and $R_{E,\nu_i}$, the luminosity is constant
while the spectral distribution of the flux still evolves due to
neutrino absorption and reemission as well as energy-exchanging
collisions.  Between $R_{L,\nu_i}$ and $R_{E,\nu_i}$ the number flux
of a given neutrino flavor need not be conserved, whereas the lepton
number flux (difference between neutrinos and antineutrinos of a given
flavor) is conserved, even for the electron flavor, because after a
few hundred milliseconds the atmosphere is in a relaxed state and does
not gain or lose lepton~number on short timescales.

Besides reactions in which energy is exchanged between neutrinos and
the stellar medium, a significant fraction of the neutrino opacity of
the stellar atmosphere is due to nearly iso-energetic neutrino-nucleon
or neutrino-nucleus scatterings; for $\nu_\mu$ and $\nu_\tau$ this
contribution in fact dominates.  This has the consequence that even
outside of $R_{E,\nu_i}$, where the flux spectrum is independent of
radius, the neutrinos still undergo many scatterings and propagate
outward by diffusion. Therefore, the local neutrino distribution
function is nearly isotropic, implying
$\langle{\mu}\rangle_{j,\nu_i}\ll 1$. Hence, it is between
$R_{E,\nu_i}$ and $R_{{\rm t},\nu_i}$, the region of a ``scattering
atmosphere'', where the naive Stefan-Boltzmann law is particularly
poor at accounting for the neutrino luminosity~\cite{Janka}.

The total opacity of electron (anti)neutrinos is dominated by
$\beta$-processes so that the distinction between the energy and
transport sphere is often not crucial---for this flavor the concept of
{\em the\/} neutrinosphere is crudely justified. For the other
flavors the distinction is crucial.

\newpage

\subsection{Pulsar Kicks by Oscillations?}
\label{sec-23}

Our description of neutrino transport reveals that there is no simple
relationship between the spectral temperature and luminosity of a
given neutrino flavor~\cite{Janka}. In numerical simulations with
nonequilibrium transport description one finds approximately equal
luminosities of neutrinos and antineutrinos of all flavors, but vastly
different spectral temperatures~\cite{Bruenn,Suzuki1,Suzuki2}.
However, the most elaborate numerical models have so far not fully
taken into account several important energy-exchange channels between
heavy-flavor neutrinos and the nuclear medium such as nucleon recoils,
(inverse) nucleon-nucleon bremsstrahlung, and collective as well as
multiple-scattering effects~\cite{Suzuki1,Hannestad}; therefore, the
spectral temperatures between the flavors are likely far more similar
than had been thought, and the difference between the energy and
transport spheres may be less pronounced.

In the limit of exact equipartition, neutrino oscillations along an
aspherical resonance surface could not produce any pulsar recoil. One
caveat is that the spectral swap between the flavors modifies their
interaction rate with the medium so that there could be a small
indirect effect. Even this possibility is diminished if the spectra
are more similar than had been thought previously.

Surely, it is not possible to calculate the pulsar recoil from a
$\nu_\tau$ Stefan-Boltzmann flux, evaluated over the aspherical
resonance surface with its varying temperature.  In typical
simulations one finds an equipartition of the total energy to within a
few percent. Any possible flux anisotropy caused by neutrino
oscillations is therefore far smaller than had been assumed in
Refs.~\cite{Kus96,Qian97,Kus97,waste,morewaste,alsowaste}.


\section{Eddington atmosphere model for flux anisotropy}
\label{sec-3}

\subsection{Description of the Model}
\label{sec-31}

In the preceding Section we have argued that in the limit of exact
energy equipartition between the emitted neutrino flavors, an
aspherical resonant oscillation surface could not cause a pulsar
recoil, except perhaps by residual higher-order effects. But the
equipartition need not be exact. This is especially true if the
oscillations are into a sterile species $\nu_s$ which would not be
emitted at all without oscillation effects~\cite{Kus97}.  
Even in this case it is
difficult to obtain a large recoil by oscillations because the
magnitude of the anisotropy does not scale with the variation of the
gas temperature along the deformed emission sphere.

An anisotropy of the neutrino emission can be established only in a
much more indirect way. The aspherical escape surface of the neutrino
flux quickly leads to a perturbation of the initial configuration by
producing an aspherical density and temperature profile of the neutron
star atmosphere. This has a feed-back effect on the temperature at the
core boundary and thus modifies the neutrino flux. Therefore, one
needs a self-consistent atmospheric model to estimate the neutrino
flux anisotropy.

\begin{figure}[t]
\epsfxsize=\hsize
\hbox to\hsize{\hss\epsfbox{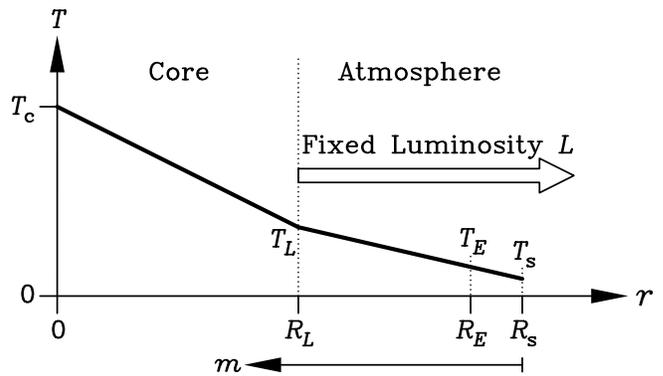}\hss}
\smallskip
\caption{\label{fig:model}
Schematic model of a nascent neutron star during the
Kelvin-Helmholtz cooling phase. The core is treated as 
one zone, the atmosphere is described self-consistently
by the Eddington atmosphere model.} 
\end{figure}

To this end we subdivide the star into the ``core'' and the
``atmosphere'' as indicated in Fig.~\ref{fig:model}. For simplicity we
take the ``luminosity sphere'' $R_L$ as the interface; outside the
neutrino luminosity is fixed. The core is characterized by a central
temperature $T_{\rm c}$, the temperature $T_L$ at the core-atmosphere
interface, and a linear temperature gradient in between.  Clearly, the
neutrino luminosity is determined by the difference between $T_{\rm
c}$ and $T_L$. In order to estimate the flux anisotropy we need to
calculate a self-consistent value for $T_L$ by matching a
self-consistent atmospheric model to the interface.

Our main task, therefore, is to construct a model for the atmosphere
which is characterized by three parameters: its mass, the temperature
$T_L$ at the bottom, and the neutrino flux $L$ which enters from below
and depends on the central temperature $T_{\rm c}$.  The primary input
quantity that varies as a function of direction is the atmospheric
mass which is given by the neutrino resonance surface.

%
%
%
%


\subsection{Neutrino Eddington Atmosphere}
\label{sec-32}

We construct a self-consistent atmospheric model by virtue of the
Eddington approximation; for neutrinos this was done in
Ref.~\cite{Schi82}, a fundamental paper that we will closely follow.
The Eddington atmosphere employs the assumption of plane-parallel
geometry, i.e.\ the atmosphere is taken to be geometrically thin
relative to the core size.  Moreover, one uses the diffusion
approximation, neutrinos and stellar medium are taken to be in thermal
equilibrium, and the neutrino and antineutrino degeneracy parameters
are taken to vanish everywhere, implying that there is no lepton
number flux through the atmosphere.  One uses neutrino and
antineutrino opacities which are equal and vary with the square of the
neutrino energy, $\Lambda_{\nu_i} = \Lambda_{\bar\nu_i} =
\Lambda_0(\epsilon^2/\epsilon_0^2)$ with $\Lambda_0$ being a constant
of dimension cm$^2{\rm g}^{-1}$ and $\epsilon_0 = {\rm const}$. This
implies that the neutrino and antineutrino phase-space distribution
functions are identical, $f_{\nu_i}= f_{\bar\nu_i}$, and can be
written in terms of the Fermi-Dirac distribution $f^{\rm eq} \equiv [1
+ {\rm e}^{\epsilon/(kT)}]^{-1}$ as
\begin{equation}
f_{\nu_i}(r,t,\mu,\epsilon)=f^{\rm eq} +
{\mu\over \Lambda}\,{\partial f^{\rm eq}\over \partial m}\,.
\label{eq-7}
\end{equation}
Here, $\mu$ is the neutrino angle cosine and $m = \int_r^{R_{\rm
s}}{\rm d}x\,\rho(x)$ the column mass density ($\rm g~cm^{-2}$) of
the atmosphere measured from the surface
inward~(Fig.~\ref{fig:model}).

These assumptions are reasonably well fulfilled for electron neutrinos
$\nu_e$ and antineutrinos $\bar\nu_e$ in the region of interest, i.e.\
between the $\nu_\mu$ and $\nu_\tau$ energy sphere $R_{E,\nu_x}$ and
those of $\nu_e$ and $\bar\nu_e$, $R_{E,\nu_e}\approx
R_{E,\bar\nu_e}$, which are very close to the neutron-star surface. On
the other hand, the thermal coupling of muon and tau neutrinos to the
stellar background ceases in the relevant regions so that we take the
atmospheric structure to be determined by the electron (anti)neutrinos
alone.  Their combined luminosity $L_{\nu_e}+L_{\bar\nu_e} = 4\pi
r^2(F_{\nu_e}+F_{\bar\nu_e})$ is assumed to be constant and given by
the inflow from the core.  In our plane-parallel model this implies
that both the area $4\pi r^2$ and the combined energy flux
$F_E=F_{\nu_e}+F_{\bar\nu_e}$ are constant.

In this model one can derive an expression for the temperature as a
function of the mass coordinate $m$ which is given by Eq.~(27) of
Ref.~\cite{Schi82} as
\begin{equation}
(kT)^2(m)=\left({9(hc)^3\over 2\pi^3\,c}\,{\Lambda_0\over 
\epsilon_0^2}\,F_E\right)m+
\left({30(hc)^3\over 7\pi^5\,c}\,F_E\right)^{1/2}.
\label{eq-8}
\end{equation}
For $m = 0$ this equation yields
the surface temperature, 
$(kT_{\rm s})^4 = F_E\, 30(hc)^3/(7\pi^5c)$,
as a function of the energy flux $F_E$. 
Using Eq.~(\ref{eq-7}) and the Rosseland mean opacity,
\begin{equation}
\Lambda_{\rm R}={7\pi^2\over 5}\,\Lambda_0\,
\left({kT\over \epsilon_0}\right)^2
\label{eq-9}
\end{equation}
(Eq.~31 in Ref.~\cite{Schi82}), one can write the energy
flux $F_E$ in terms of the derivative of the neutrino and
antineutrino energy density $\varepsilon\equiv\varepsilon_{\nu_e}
+\varepsilon_{\bar\nu_e}$ as
\begin{equation}
F_E={c\over 3\Lambda_{\rm R}}\,{\partial\varepsilon\over
\partial m}
\label{eq-10}
\end{equation}
(Eq.~28 in Ref.\cite{Schi82}).

Equations~(\ref{eq-9}) and (\ref{eq-10}) can be used to relate the
energy flux coming from the core to the temperature $T_L$ at the
core-atmosphere interface.  Taking Eq.~(\ref{eq-3.1}) for the neutrino
energy density with $2{\cal F}_3(0) = 7\pi^4/60$~\cite{Blud78} and
$m_{\rm c}$ as the column mass density of the core, and evaluating
Eq.~(\ref{eq-9}) in the one-zone approximation
for an average core temperature, $T^2\equiv {1\over
2}(T_{\rm c}^2+T_L^2)$ where $T_{\rm c}$ is the temperature at the
center, one finds
\begin{equation}
F_E={2\pi^3\over 9}\,{c\,\epsilon_0^2\over
\Lambda_0(hc)^3\,m_{\rm c}}\,
\left[(kT_{\rm c})^2-(kT_L)^2\right]\,.
\label{eq-11}
\end{equation}
When one defines the column mass density of the atmosphere between
radius $R_L$ and the surface at $R_{\rm s}$ as $m_L\equiv
\int_{R_L}^{R_{\rm s}}{\rm d}r\,\rho(r)$ and plugs Eq.~(\ref{eq-11})
into Eq.~(\ref{eq-8}), one ends up with
\begin{eqnarray}
(kT_L)^2&=&\alpha\,\left[(kT_{\rm c})^2-(kT_L)^2\right]
\nonumber\\
\noalign{\smallskip}
&+&\sqrt{\beta\left[(kT_{\rm c})^2-(kT_L)^2\right]}
\label{eq-12}
\end{eqnarray}
where
\begin{eqnarray}
\alpha\, &\equiv&\, {m_L\over m_{\rm c}}\,, \nonumber\\
\beta\, &\equiv&\, {20\over 21\pi^2}\,{\epsilon_0^2\over 
\Lambda_0 m_{\rm c}}= {4\over 3}\,{(kT_{\rm c})^2\over
\Lambda_{\rm R}(T_{\rm c})\,m_{\rm c}}\,.
\label{eq-13}
\end{eqnarray}
The second expression for $\beta$ was derived by using
Eq.~(\ref{eq-9}) with $T = T_{\rm c}$. The product $\Lambda_{\rm
R}(T_{\rm c})\,m_{\rm c}$ is a measure of the optical depth of the
core for $\nu_e$ and $\bar\nu_e$ with the Rosseland mean free path
being computed for a neutrino spectrum with temperature $T_{\rm c}$.
Equation~(\ref{eq-12}) can be solved for~$T_L^2$,
\begin{eqnarray}
(kT_L)^2&=&\frac{1}{2(1+\alpha)^2}\biggl(
2\alpha(1+\alpha)(kT_{\rm c})^2 - \beta\nonumber\\
&&\hskip4.5em
{}\pm\sqrt{\beta\left[\beta+4(1+\alpha)(kT_{\rm c})^2\right]}
\biggr)\,.
\label{eq-14}
\end{eqnarray}
Since $\alpha\ll 1$ and $\beta\ll (kT_{\rm c})^2$ (see below), 
this result can be approximated to first order in $\alpha$ by
\begin{equation}
kT_L\,\cong\,\sqrt{\alpha}\,kT_{\rm c}\,.
\label{eq-15}
\end{equation}
This and Eqs.~(\ref{eq-11}) and~(\ref{eq-13}) yield
\begin{eqnarray}
F_E&=&{7\pi^5\,c\over 30(hc)^3}\,
\beta\,\left[(kT_{\rm c})^2
-(kT_L)^2\right]\nonumber\\
&\cong&{7\pi^5\,c\over 30(hc)^3}\,\beta(1-\alpha)\,
(kT_{\rm c})^2 \,.
\label{eq-16}
\end{eqnarray}

For conditions representative of the phase where the nascent neutron
star loses most of its binding energy by neutrino emission, the
central core temperature is around $kT_{\rm c}\approx 30$--$50\,$MeV
and the temperature at the base of the atmosphere (near or somewhat
inside the muon and tau neutrino energy sphere) is $kT_L\approx
10$--$15\,$MeV.  Equation~(\ref{eq-15}) thus implies $\alpha\approx
1/10$ as a typical number. The optical depth of the core for $\nu_e$
and $\bar\nu_e$ is of the order $\Lambda_{\rm R}(T_{\rm c})m_{\rm
c}\sim 10^5$, so that $\beta\sim 10^{-5}(kT_{\rm c})^2$. Using these
numbers in Eq.~(\ref{eq-16}) one gets a luminosity $L_E = 4\pi R_{\rm
s}^2 F_E\sim 2\times 10^{51}\, (R_{\rm s}/10\,{\rm km})^2(kT_{\rm
c}/50\,{\rm MeV})^4\,{\rm erg~s}^{-1}$ for $\nu_e$ plus $\bar\nu_e$.
Assuming all neutrino and antineutrino flavors contribute equally,
this corresponds to a total neutrino luminosity of $L_{\nu}\sim
6\times 10^{51}\,{\rm erg~s}^{-1}$, in good agreement with detailed
numerical models.


\subsection{Emission Anisotropy}
\label{sec-33}

In order to estimate the anisotropy of the neutrino emission we take
the deformed effective neutrinosphere to be what corresponds to the
atmospheric surface $R_{\rm s}$ in the previous section. The column
mass density of the atmosphere, i.e.\ between the core boundary $R_L$
and $R_{\rm s}$, is taken as
\begin{eqnarray}
m_L&=&m_{L0} + \delta m_L\cos\phi\nonumber\\ 
&=&{m_{+}+m_{-}\over 2} + {m_{+}-m_{-}\over 2}\,\cos\phi
\label{eq-17}
\end{eqnarray}
where $\cos\phi = ({\bf q}\cdot{\bf B})/q$ is the cosine of the angle
of the neutrino momentum ${\bf q}$ relative to the direction of the
magnetic field and $m_\pm$ are the atmospheric column densities that
correspond to $\cos\phi=\pm 1$.  With Eq.~(\ref{eq-17}) inserted into
Eq.~(\ref{eq-13}) and Eq.~(\ref{eq-16}) one obtains
\begin{equation}
F_{\rm E}(\cos\phi)\propto\beta\,(kT_{\rm c})^2\,
\left(1-{m_{L0}\over m_{\rm c}} - 
{\delta m_L\over m_{\rm c}}\,\cos\phi\right). 
\label{eq-18}
\end{equation}
The asymmetry in the third component of the neutrino momentum can now
be estimated as
\begin{eqnarray}
{\Delta q\over q}&\approx&{1\over 6}\,
{\int_{-1}^{+1}{\rm d}\cos\phi\,\cos\phi\,
F_{\rm E}(\cos\phi)\over \int_{-1}^{+1}{\rm d}\cos\phi\,
F_{\rm E}(\cos\phi)}\nonumber\\
&\approx&-\,{1\over 18}\,{\delta m_L\over m_{\rm c}}\,.
\label{eq-19}
\end{eqnarray}
This result is accurate to first order in $\alpha_0 \equiv
m_{L0}/m_{\rm c}\ll 1$ and we have assumed, as in Refs.~\cite{Kus96}
and~\cite{Qian97}, that only one neutrino species is responsible for
the anisotropy which carries off about $1/6$ of the total energy.

The mass difference $\delta m_L$ is connected with
the radial deformation $\delta r$ of the surface of resonance,
which is defined by $r(\phi) = r_0 + \delta r\,\cos\phi$, through
\begin{equation}
\delta m_L={1\over 2}(m_{+}-m_{-})=\rho_0\,\delta r\,.
\label{eq-20}
\end{equation}
Here, $\rho_0$ is the mean density at the surface of resonance. For
the width $\delta r$ in dependence of the strength of the magnetic
field $B$ one finds~\cite{Kus96,Qian97}
\begin{equation}
\delta r={eB\over 2}\,\left({3n_e\over \pi^4}\right)^{\! 1/3}
\left({{\rm d}n_e\over {\rm d}r}\right)^{\! -1}\approx
{3\over 2}\,{eB\over \psi_e^2}\,h_{n_e}
\label{eq-21}
\end{equation}
where $h_{n_e} = |\partial\ln n_e/\partial r|_{r_0}^{-1}$ is the scale
height for changes of the electron number density near $r_0$ and
$\psi_e\approx hc(3n_e/8\pi)^{1/3}$ is the chemical potential of the
electrons. With the definition
\begin{equation}
\gamma\equiv{3\over 2}\,{eB\over \psi_e^2}\,\approx\,
0.22\,\left(20\,{\rm MeV}\over \psi_e\right)^{\! 2}\,
\left({B\over 10^{16}\,{\rm G}}\right)
\label{eq-22}
\end{equation}
and the density scale height $h_{\rho} = |\partial\ln \rho/\partial
r|_{r_0}^{-1}$ near ${r_0}$, Eqs.~(\ref{eq-20}) and (\ref{eq-21})
yield
\begin{equation}
\delta m_L=\rho_0 h_{\rho}\,\gamma\,{h_{n_e}\over
h_{\rho}}\,.
\label{eq-23}
\end{equation}
Finally, with Eq.~(\ref{eq-19}) one ends up with
\begin{equation}
{\Delta q\over q}\approx -{1\over 18}\,{\rho_0 h_{\rho}\over
m_{\rm c}}\,\gamma\,{h_{n_e}\over h_{\rho}}\,.
\label{eq-24}
\end{equation}
Taking $\rho_0 h_{\rho}/m_{\rm c} < m_{L0}/m_{\rm c} \approx 1/10$
and $h_{n_e}/h_{\rho} \alt 1$~\cite{Qian97} leads to the numerical
estimate
\begin{equation}
{\Delta q\over q}< -0.0012\,
\left({20\,{\rm MeV}\over \psi_e}\right)^{\! 2}\,
\left({B\over 10^{16}\,{\rm G}}\right) \,.
\label{eq-25}
\end{equation}
This result is at least 10 times smaller than the an\-iso\-tropy
derived in Ref.~\cite{Qian97}. 

Therefore, the kick mechanism based on a deformation of the effective
neutrinosphere requires more than an order of magnitude larger magnetic
fields than estimated in Refs.~\cite{Qian97,waste,morewaste} whose 
analysis already reduced the effect originally discussed by Kusenko and
Segr\`e~\cite{Kus96}. For a neutrino emission anisotropy of 1\%,
corresponding to a recoil velocity of the nascent neutron star of 
approximately 300$\,{\rm km\,s}^{-1}$, one needs magnetic fields
in excess of about $10^{17}\,$G near the stellar surface.

The value of $\gamma$ in Eq.~(\ref{eq-22}) is sensitive to the
electron chemical potential. In Ref.~\cite{Qian97} $\gamma$ was 
evaluated by using $Y_e=n_e/n_b\approx 0.1$ for the electron fraction 
($n_b$ is the baryon number density) at a density $\rho\approx
10^{12}\,{\rm g\,cm}^{-3}$. A discussion of the uncertainties of
this choice can be found in Appendix~A where the structure of the 
protoneutron star atmosphere is self-consistently determined 
from a simple, analytical model. Typically, $\gamma$ decreases
during the neutrino cooling of the nascent neutron star because 
the atmosphere becomes denser and more compact as the star cools
and deleptonizes. This disfavors large $\delta r$ at intermediate
and late times during the Kelvin-Helmholtz cooling when most of the
gravitational binding energy of the neutron star is emitted in
neutrinos. This makes large emission anisotropies even more unlikely.

 
\section{Summary and discussion}
\label{sec-4}

We have argued that the neutrino-oscillation scenarios for neutron
star kicks~\cite{Kus96,Qian97,Kus97,waste,morewaste,alsowaste} 
suffer from two
serious flaws; both problems are related to an incorrect picture of
neutrino transport in the atmosphere of a protoneutron star.

First, when the neutrino luminosity is equipartitioned between the
flavors, no significant recoil can be produced because only an
indirect, higher-order effect remains which is associated with the
spectral swap of two neutrino flavors. Spectral differences imply a
change of the neutrino interaction with the stellar background and
thus affect the neutrino transport from the core to the surface
through a modified atmospheric temperature profile. While we cannot
estimate the magnitude of the resulting small kick velocity, we are
convinced that it is a very small effect.  Moreover, current supernova
models overestimate the spectral differences between $\nu_e$ and
$\nu_{\tau}$ because neutrino interactions have not been taken into
account which enhance the thermal coupling between $\nu_{\tau}$ and
the stellar medium~\cite{Suzuki2,Hannestad}.

Second, when the luminosities are taken to be vastly different (as had
effectively been assumed in previous papers), again no effect obtains
in zeroth order because a shift of the location of the neutrinosphere
does not affect the neutrino luminosity. The latter is governed by the
core emission, not by local processes in the atmosphere.

However, the atmosphere adjusts to the oscillation-induced
modification of its transport capabilities, causing a small,
higher-order effect due to an altered temperature gradient in the
core; in our treatment it was expressed as a change of the temperature
at the core-atmosphere interface.  In this sense the neutrino flux
from the core fixes the atmospheric temperature profile and,
conversely, the atmosphere determines the core emission. In contrast,
in Refs.~\cite{Kus96,Qian97,Kus97,waste,morewaste,alsowaste} it had
been assumed that the unperturbed atmospheric temperature is a measure
of the neutrino luminosity by virtue of the Stefan-Boltzmann law.

Both of our arguments imply a huge suppression of the pulsar recoils
calculated in
Refs.~\cite{Kus96,Qian97,Kus97,waste,morewaste,alsowaste}, but a
realistic quantitative estimate of the residual effects is not
possible with our simple analytic tools. A detailed numerical
treatment would be extremely difficult, and the motivation for such an
effort is minimal because most likely one would confirm what now looks
like a non-effect.  In any case, it is clear that the oscillation
scenarios require much larger magnetic fields than had been
contemplated in
Refs.~\cite{Kus96,Qian97,Kus97,waste,morewaste,alsowaste} and thus
probably take one beyond what is astrophysically motivated.

We find it disappointing that both the cumulative parity violation and
the neutrino oscillation scenario, which seemed to work with
reasonable magnetic field strengths, do not survive a self-consistent
discussion.  Of course, it remains possible that huge magnetic fields
($\agt 10^{16}\,$G) with an asymmetric distribution in the core of the
protoneutron star cause sufficiently asymmetric neutrino
opacities for a large neutrino rocket effect~\cite{Bisno93}.  It is
also possible that asymmetric neutrino emission has nothing to do with
the pulsar kicks. Either way, it does not look as if the pulsar
velocities can be attributed to neutrino oscillations within presently
discussed scenarios.

{\bf Note Added in Proof.}---In a recent preprint~[21], Kusenko and
Segr\`e criticize our analysis and affirm their previous results~[5].
Their main objection against our work is that allegedly we ignored
neutrino absorption via charged-current interactions and assumed equal
opacities for all neutrino flavors.  However, opacity differences
between electron neutrinos and muon/tau neutrinos {\it were}, of
course, included in our analysis.  In our analytical model, the
opacities determine the column density of the atmosphere between the
core boundary and the neutron-star ``surface''. The latter was taken 
to be the effective sphere of neutrino-matter decoupling which is 
located at different radii for the different neutrino flavors.

In their new analysis~[21], Kusenko \& Segr\`e estimate the neutron
star kick associated with anisotropic resonant flavor conversions by
considering the asymmetric absorption of electron neutrinos in the
magnetized neutron star atmosphere. This approach is based on the same
assumptions as their previous one~[5] and therefore it is not
astonishing that their original estimate of the magnitude of the 
pulsar kick is confirmed. However, resonant flavor conversions in the
neutron star atmosphere cannot cause a persistent emission or absorption
anisotropy because the neutrino
luminosity is determined by the flux from the core, and the atmosphere
adjusts to the inflow from below within a time which is very short
compared to the neutrino-cooling time of the nascent neutron
star. Therefore, the absorption anisotropy calculated by Kusenko and
Segr\`e is a transient phenomenon until the enhanced absorption is
balanced by the reemission of neutrinos and the atmosphere has
approached a new stationary state.  Of course, if the magnetic field
is initially present rather than being ``switched on,'' even this
transient phenomenon will not occur---the unperturbed atmosphere
simply never exists.

As stressed in the main text of our paper, the large kick velocities
found by Kusenko and Segr\`e are an artifact of using an unperturbed
atmospheric model instead of a self-consistent one. Accepting that a
young neutron star is well described by our core-atmosphere picture
(and Kusenko and Segr\`e do not seem to question this crucial premise
of our work), our conclusion that there is no zeroth-order kick 
caused by neutrino oscillations in the atmosphere is
rigorous and as such not subject to debate. As described in our paper,
there will be a higher-order effect due to a modification of the
temperature at the core-atmosphere interface caused by the asymmetry
of the self-consistent atmospheric structure.  The resulting kick is
much smaller than estimated by Kusenko and Segr\`e. Our discussion,
however, does not exclude that a larger, globally asymmetric neutrino
emission may develop if the anisotropies are produced by effects in
the dense inner core of the neutron star.


\section*{Acknowledgments}

This work was supported, in part, by the Deutsche
Forschungsgemeinschaft under grant No.\ SFB-375.


\appendix

\section{Simple Model for Protoneutron~Star~Atmosphere}

A crucial parameter for estimating $\delta r$ in Eq.~(\ref{eq-21}) is
the electron number density $n_e$ in the protoneutron star atmosphere
between the energy spheres of $\nu_e$ and heavy-flavor neutrinos 
$\nu_{\mu}$ and $\nu_{\tau}$.  Making use of the
fact that the degeneracy parameter of electron neutrinos approaches
zero, $\eta_{\nu_e}=\eta_e+\eta_p-\eta_n\to 0$, as the protoneutron
star atmosphere deleptonizes and the electron chemical potential
$\psi_e=kT\eta_e$ decreases~\cite{Suzuki2,PNS}, 
one can easily estimate $n_e$ and the
electron number fraction $Y_e = n_e/n_b = (n_{e^-}-n_{e^+})/n_b$ where
$n_b = \rho/m_u$ is the number density of baryons and $m_u$ the atomic
mass unit. With $Y_e = Y_p$, $Y_n = 1-Y_p$, and $\eta_n-\eta_p \approx
\ln(Y_n/Y_p)$ for Boltzmann gases of neutrons $n$ and protons $p$ with
$m_n\approx m_p$, one finds as a very good approximation
\begin{eqnarray}
Y_e&\approx&{8\pi m_u (kT)^3 \over (hc)^3 \rho}\, 
{\eta_e\over 3}\,\left(\pi^2+\eta_e^2\right) \nonumber\\
&\approx&9.1\times 10^{-3}\, 
\left({kT\over 5~{\rm MeV}}\right)^{\! 3}
\,\left({\rho\over 10^{11}~{\rm g~cm}^{-3}}\right)^{\! -1}
 \nonumber\\
&&{}\times\ln\left({1\over Y_e}-1\right)\,
\Biggl\lbrace\pi^2+
\left[\ln\left({1\over Y_e}-1\right)\right]^2\Biggr\rbrace\,.
\label{eq-26}
\end{eqnarray}
This equation shows that $Y_e$ decreases with increasing density
$\rho$ for a given temperature.

The temperature as a function of column mass density $m$ is given by
Eq.~(\ref{eq-8}) as $(kT)^2 = Am+(kT_{\rm s})^2$. In order to
determine $\rho(m)$, we employ hydrostatic equilibrium which yields
for the pressure $P = gm+P_{\rm s}$ as a function of $m$.  Here
$P_{\rm s}$ is the pressure at the surface and $g = GM_{\rm ns}/R_{\rm
s}^2$ is the gravitational acceleration near the surface of the
protoneutron star which is nearly constant in the thin, plane-parallel
atmosphere. Since the layers between the energy spheres of electron
neutrinos and muon and tau neutrinos are dense ($\rho\sim
10^{11}$--$10^{14}~{\rm g~cm}^{-3}$) and rather cool ($kT\sim
3$--$10~{\rm MeV}$), the pressure is dominated by baryons (see Fig.~8
of Ref.~\cite{Woos86}) so that we can take $P \approx kT\rho/m_u$ to
obtain
\begin{equation}
\rho(m)\approx{(gm+P_{\rm s})m_u\over kT}\,.
\label{eq-27}
\end{equation}
With Eqs.~(\ref{eq-8}) and~(\ref{eq-27}) one derives
\begin{equation}
{(kT)^3\over \rho}\approx{\left[{Am+(kT_{\rm s})^2}\right]^2\over
m_u gm+ \rho_{\rm s}kT_{\rm s}}\,.
\label{eq-28}
\end{equation}
From this relation and Eq.~(\ref{eq-26}) one can see that $Y_e$ first
decreases only slightly, then increases with rising $m$, i.e.\ on the
way inward into the atmosphere. Therefore, the minimum value of the
electron fraction can be found very close to the electron neutrino
energy sphere $R_{E,\nu_e}\approx R_{\rm s}$. This confirms
$h_{n_e}/h_{\rho}\alt 1$ because $\rho$ as well as $Y_e$ have a
negative gradient interior to $R_{E,\nu_e}$. Typical conditions near
$R_{E,\nu_e}$ at early times during the protoneutron star evolution
are~\cite{Suzuki2,PNS} $kT_{\rm s}\approx 3~$MeV and $\rho_{\rm
s}\approx 10^{11}~{\rm g~cm}^{-3}$ which yields $Y_e(R_{\rm s})
\approx 0.078$. At later times the temperature in the atmosphere drops
due to cooling~\cite{Suzuki2,PNS}, and according to Eq.~(\ref{eq-27})
the density in the atmosphere must become higher and the density
gradient steeper. Therefore, the electron neutrinosphere, which is
assumed to be located at a certain value of the optical depth, moves
to higher densities. Typical conditions then are $kT_{\rm s} \approx
5~$MeV and $\rho_{\rm s} \approx 10^{12}~{\rm g~cm}^{-3}$ for which
one gets $Y_e(R_{\rm s}) \approx 0.050$. Even later, one has $kT_{\rm
s} \approx 5~$MeV and $\rho_{\rm s} \approx 5\times 10^{13}~{\rm
g~cm}^{-3}$ which gives $Y_e(R_{\rm s}) \approx 0.004$.

Although $Y_e(R_{\rm s})$ decreases as the protoneutron star cooling
goes on, the number density $n_e(R_{\rm s})$ nevertheless increases
because of the rising density $\rho_{\rm s}$. In the listed examples,
$n_e(R_{\rm s})$ changes from $4.7\times 10^{33}~{\rm cm}^{-3}$
through $3.0\times 10^{34}~{\rm cm}^{-3}$ to $1.2\times 10^{35}~{\rm
cm}^{-3}$. From Eq.~(\ref{eq-21}) we conclude that this disfavors
large $\delta r$ at intermediate and late times during the
Kelvin-Helmholtz neutrino cooling of the nascent neutron star.


\end{document}